\input harvmac 
\input epsf.tex
\def\IN{\relax{\rm I\kern-.18em N}} 
\def\IR{
\relax{\rm I\kern-.18em R}} \font\cmss=cmss10 
\font\cmsss=cmss10 at 7pt \def\IZ{\relax\ifmmode\mathchoice 
{\hbox{\cmss Z\kern-.4em Z}}{\hbox{\cmss Z\kern-.4em Z}} 
{\lower.9pt\hbox{\cmsss Z\kern-.4em Z}} {\lower1.2pt\hbox{
\cmsss Z\kern-.4em Z}}
\else{\cmss Z\kern-.4em Z}\fi} 

\overfullrule=0mm

\newcount\figno \figno=0
\newcount\figtotno      
\figtotno=0
\newdimen\captionindent 
\captionindent=1cm 
 
\newcount\figno
\figno=0
\def\fig#1#2#3{ \par\begingroup\parindent=0pt
\leftskip=1cm\rightskip=1cm\parindent =0pt 
\baselineskip=11pt
\global\advance\figno by 1
\midinsert
\epsfxsize=#3
\centerline{\epsfbox{#2}}
\vskip 12pt
{\bf Fig. \the\figno:} #1\par
\endinsert\endgroup\par
}
\def\figlabel#1{\xdef#1{\the\figno}} 
\def\encadremath#1{\vbox{\hrule\hbox{\vrule\kern8pt 
\vbox{\kern8pt \hbox{$\displaystyle #1$}\kern8pt} 
\kern8pt\vrule}\hrule}} \def\enca#1{\vbox{\hrule\hbox{
\vrule\kern8pt\vbox{\kern8pt \hbox{$\displaystyle #1$}
\kern8pt} \kern8pt\vrule}\hrule}}
\def\tvi{\vrule height 12pt depth 6pt width 0pt} 
\def\tv{\tvi\vrule}

\def\IR{\relax{\rm I\kern-.18em R}}
\font\cmss=cmss10 \font\cmsss=cmss10 at 7pt 
\def\IZ{\relax\ifmmode\mathchoice
{\hbox{\cmss Z\kern-.4em Z}}{\hbox{\cmss Z\kern-.4em Z}} 
{\lower.9pt\hbox{\cmsss Z\kern-.4em Z}}
{\lower1.2pt\hbox{\cmsss Z\kern-.4em Z}} 
\else{\cmss Z\kern-.4em Z}\fi} \def\buildrel#1\under#2{ 
\mathrel{\mathop{\kern0pt #2}\limits_{#1}}}

%%%%%%%%%%%%%%%%%%%%%%%%%%%%%%%%%%%%%%%%%%%%% 
\Title{UNC-CH-MATH-98/2}
{{\vbox {
\centerline{Folding Transitions of the Square-Diagonal Lattice}}}}
\bigskip
\centerline{P. Di Francesco\footnote*{e-mail: philippe@math.unc.edu},}
\bigskip
\centerline{\it Department of Mathematics,} 
\centerline{\it University of North Carolina at Chapel Hill,} 
\centerline{\it  CHAPEL HILL, N.C. 27599-3250, U.S.A.} 
\vskip .5in
%abstract
\noindent 
We address the problem of "phantom" folding of the tethered membrane
modelled by the two-dimensional square lattice, with bonds on the edges 
and diagonals of each face. 
Introducing bending rigidities $K_1$ and $K_2$ for respectively long and 
short bonds, we derive the complete phase diagram of the model, using transfer
matrix calculations. The latter displays two transition curves, one
corresponding to a first order (ferromagnetic) folding transition, and the
other to a continuous (anti-ferromagnetic) unfolding transition. 

\Date{04/98 $\ \ $ PACS: 64.60.-i $\ \ \ \ \ $
Keywords: polymerized membrane, folding, vertex model, phase transition}  
%\draft
%\writetoc

%references

%\nref\KN{Y. Kantor and D.R. Nelson, {\it Crumpling Transition 
%in Polymerized Membranes}, Phys. Rev. Lett. {\bf 58} (1987) 2774
%and {\it Phase Transitions in Flexible Polymeric Surfaces}, 
%Phys. Rev.  {\bf A 36} (1987) 4020.}
%\nref\NP{D.R. Nelson and L. Peliti, {\it Fluctuations in Membranes
%with Crystalline and Hexatic Order}, J. Physique {\bf 48} (1987) 1085.}
%\nref\PKM{M. Paczuski, M. Kardar and D.R. Nelson, {\it Landau
%Theory of The Crumpling Transition}, Phys. Rev. Lett. 
%{\bf 60} (1988) 2638.}
%\nref\DG{F. David and E. Guitter, {\it Crumpling Transition
%in Elastic Membranes: Renormalization Group Treatment},
%Europhys. Lett. {\bf 5} (1988) 709.}
%\nref\BESP{M. Baig, D. Espriu and J. Wheater, {\it Phase Transitions in 
%Random Surfaces}, Nucl. Phys. {\bf B314} (1989) 587;
%R. Renken and J. Kogut, {\it Scaling Behavior at the Crumpling Transition},
%Nucl. Phys. {\bf B342} (1990) 753; R. Harnish 
%and J. Wheater, {\it The Crumpling Transition of Crystalline Random Surfaces},
%Nucl. Phys. {\bf B350} (1991) 861; J. Wheater and P. Stephenson,
%{\it On the Crumpling Transition in Crystalline Random Surfaces},
%Phys. Lett. {\bf B302} (1993) 447.}
\nref\MEA{P. Di Francesco, O. Golinelli and E. Guitter, 
{\it Meander, Folding
and Arch Configurations}, Mathl. Comput. Modelling, 
Vol. {\bf 26}, No.8-10 (1997) 97-147, 
{\it Meanders and the Temperley-Lieb Algebra},
Commun. Math. Phys. {\bf 186} (1997), 1-59
and {\it Meanders: a direct enumeration approach},
Nucl. Phys. {\bf B482[FS]} (1996), 497-535; P. Di Francesco, 
{\it Meander Determinants},
Commun. Math. Phys. {\bf 191} (1998) 543-583.}
\nref\KAN{Y. Kantor and M.V. Jari\'c, Europhys. Lett. {\bf 11} (1990) 157.}
\nref\DGE{P. Di Francesco and E. Guitter {\it Entropy of Folding of 
the Triangular Lattice}, Europhys. Lett. {\bf 26} (1994) 455.}
\nref\DGT{P. Di Francesco and E. Guitter {\it Folding Transition of the
Triangular Lattice}, Phys. Rev. {\bf E50} (1994) 4418-4426.}
\nref\PDF{P. Di Francesco {\it Folding the Square-Diagonal Lattice},
preprint UNC-CH-MATH 98/1, cond-mat/9803051, accepted for publication
in Nucl. Phys. {\bf B} (1998).} 
%\nref\DGGF{M. Bowick, P. Di Francesco, O. Golinelli and E. Guitter 
%{\it 3D Folding of the triangular lattice}, Nucl. Phys. 
%{\bf B450[FS]} (1995) 463-494.}
%\nref\TLA{H. Temperley and E. Lieb, {\it Relations between the Percolation
%and Coloring Problems and other Graph-Theoretical Problems associated with
%regular Planar Lattices: Some Exact Results for the Percolation
%Problem}, Proc. Roy. Soc. {\bf A322} (1971) 251-280; see also the book
%by P. Martin, {\it Potts Models and Related Problems in Statistical
%Mechanics}, World Scientific, Singapore (1991) for a review.}
%\nref\BAX{R.J. Baxter, {\it Exactly Solved Models in Statistical
%Mechanics}, Academic Press, London (1982).}
%\nref\DEG{P. Di Francesco, B. Eynard and E. Guitter, {\it Coloring
%Random Triangulations}, cond-mat/9711050, to appear in 
%Nucl. Phys. {\bf B} (1998).}

%text
\newsec{Introduction}
\par

Models for discrete polymerized membranes with rigid bonds 
display interesting physical behaviors, as the only way to
change their spatial configuration is through folding.
We will consider here only "phantom" folding, in which the
membrane is allowed to interpenetrate itself, as opposed to
self-avoiding folding, which introduces considerable
mathematical difficulties, as was already noticed in the
case of compact (one-dimensional) polymer folding \MEA.

The entropy of folding  of a regular triangular lattice has been 
first studied numerically in \KAN\ and later computed exactly in \DGE,
by mapping the model onto the edge tri-coloring problem of the 
triangular lattice. The next step was the introduction of a bending
rigidity, namely an energy which favors either folded edge states or
unfolded ones (typically, one associates the Boltzmann weights
$e^{\pm K}$ respectively to each unfolded or folded edge). 
Numerical studies \DGT\ have shown the existence of a
first order folding transition between completely folded states
(with a certain entropy) and a unique completely flat state, at a 
critical value $K=K_c>0$. 
On the other hand, the model also displays a continuous 
unfolding transition
between a completely folded state and a partially unfolded one at
another critical value $K=K_*<0$. Only little numerical
evidence for the latter 
transition was found in \DGT, but it was argued that for $K<0$,
starting from a completely folded state, edges could be unfolded 
along loops, a phenomenon typical of continuous (Ising-like) transitions.

In the present paper, we carry out an analogous study for the
square-diagonal lattice folding problem introduced in \PDF. The
square-diagonal lattice (see Fig.1 below) has two types of edges, 
long and short, it is therefore natural to attach to them two different
types of Boltzmann weights, $e^{\pm K_i}$, $i=1,2$ for respectively short and
long edges. In addition to getting better numerical estimates when 
$K_2=0$ or $K_1=0$, we will be able to investigate the full critical
structure of the model in the plane $(K_1,K_2)$.

\medskip
The paper is organized as follows.
In Sect.2, we review the 2-dimensional folding problem of the 
square-diagonal lattice, and its reformulation as an edge 
tangent-vector model, and various face/vertex models. 
In the presence of bending rigidity, the transfer matrix of the
model is simple enough to permit numerical estimates of the free energy and
magnetization, by extraction of its largest eigenvalue.
Sect.3 is devoted to the numerical study of the rigid short edge model, 
with $K_2=0$,
in which only the short edges have a bending rigidity, whereas Sect.4
deals with the rigid long edge model, with $K_1=0$. Both models
confirm the first order folding transition of \DGT, at some critical 
values $K=K_{c,i}>0$.
In Sect.5, the two above models are studied for $K_i<0$, and better evidence
is found for a continuous unfolding transition at some critical values 
$K_i=K_{i,*} <0$.  
The general rigid long and short edge model is studied in Sect.6,
where we find the complete phase diagram of the model, including two 
critical curves corresponding to the folding and unfolding transitions, 
and which are separated by the line $K_2=-2K_1$.
\par
\newsec{Folding the Square-Diagonal Lattice}
\par
\fig{The Square-Diagonal lattice. It has two types of vertices,
respectively 4- and 8-valent, and two types of edges, short
(length $1$) and long (length $\sqrt{2}$).}{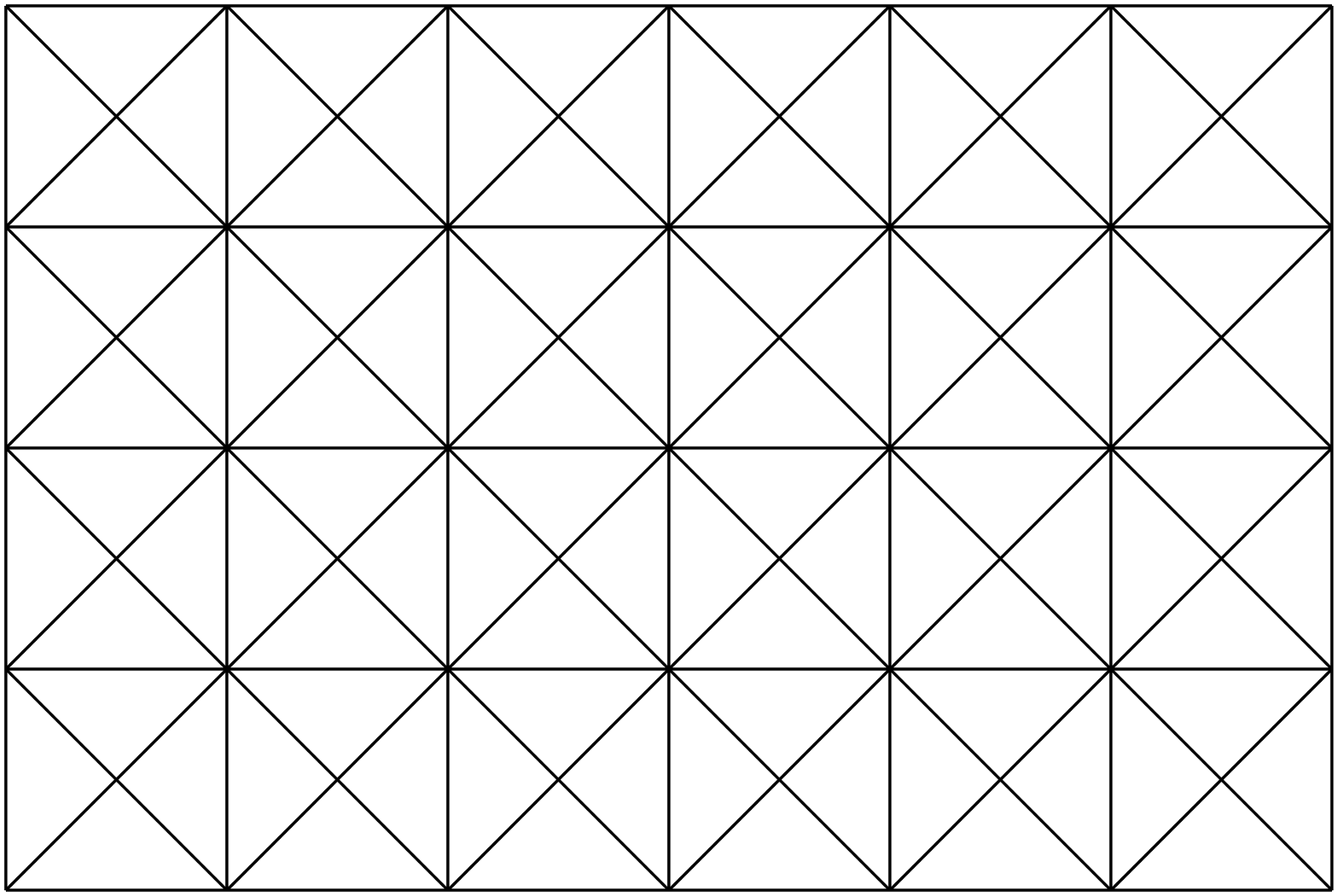}{6.cm}
\figlabel\sqdiaglat
\par
Let us consider the {\it Square-Diagonal lattice}, obtained from 
the standard square lattice by drawing the two diagonals 
on each face, as depicted in Fig.\sqdiaglat.  Each face of 
this lattice
is a triangle with two "short" edges of length $1$ and one
"long" edge of length $\sqrt{2}$. In this paper we study the
possible foldings of the lattice, in which the edges serve
as hinges between adjacent faces. The simplest such foldings
are those with a two-dimensional final state, namely the foldings
of the lattice into itself. Here we consider only {\it phantom}
folding, in which the lattice may interpenetrate itself, and
distinguish only between distinct final two-dimensional (folded)
states of the lattice, namely its {\it folding configurations}.

In \DGT, we have shown that this two-dimensional folding problem
is equivalent to various vertex models. We first defined the
tangent vectors to the lattice as the vectors $\vec{t}$ drawn 
along its edges, oriented in a compatible way, so that the sum 
\eqn\rulfac{ \sum_{\rm face} \ \vec{t} ~=~ \vec{0} }
taken around each face of the lattice vanishes.  The short tangent
vectors may take the 4 values $\pm \vec{e_1}$, $\pm\vec{e_2}$,
where $(\vec{e_1},\vec{e_2})$ is an orthonormal basis for 
the short edges, whereas the long edges may take the 4 values 
$\pm \vec{e_1}\pm\vec{e_2}$.

A folding configuration of the lattice is entirely determined by 
the images of these tangent vectors under a continuous,
length-preserving folding map $\rho$, 
with the constraint
\eqn\consrho{ \sum_{\rm face} \ \rho(\vec{t})~=~ \vec{0} }
around each face of the lattice. Note that the images of the tangent 
vectors under $\rho$ may take only values in the corresponding 
abovementioned sets of 4.

By focusing on either short or long edges (the images of either class
are sufficient to reconstruct all other images, thanks to the 
rule \consrho), we may consider the folding configurations as either 
short or long-edge configurations on the lattice, with constraints
inherited from the rule \consrho. In both cases,
we end up with statistical models with 4 edge variables, and only
finitely many allowed vertices.

\fig{The 28 possible face configurations of the rigid short edge model.
The variable $a$ may take any value mod 4. We have indicated the
Boltzmann weights including a bending energy $K_1$ per
short edge and a magnetic field $h$.}{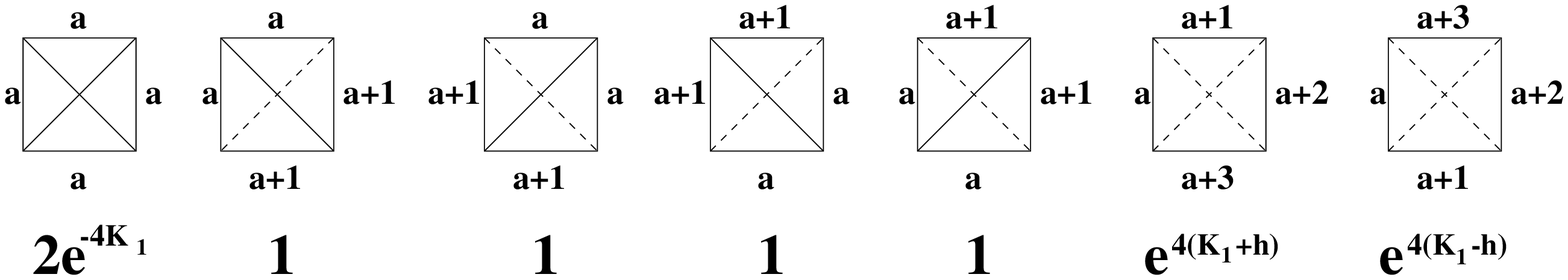}{12cm}
\figlabel\folon

On long edges, the constraint \consrho\ amounts to relating the
four long edge images around each square face made of 4 triangles.
Let us first introduce the orthogonal basis $(\vec{f_1},\vec{f_2})=
(\vec{e_1}+\vec{e_2},\vec{e_2}-\vec{e_1})$ for long edge vectors,
and label the edge values $\vec{f_1},\vec{f_2},-\vec{f_1},-\vec{f_2}$
respectively by $0,1,2,3$. In terms of these 
values, the only edge configurations around a 
square face which are allowed by \consrho\ are the
28 cases depicted in Fig.\folon.
They correspond respectively to the following folding states of the 
inner short edges: all four folded (4), two folded along the same diagonal
(16), none folded (8). In Fig.\folon,
we have represented in dashed lines the unfolded short edges, 
and in solid lines the folded ones.  Introducing a bending energy
$J_1$ to distinguish between folded and unfolded short edges,
we are led to weight the first case of Fig.\folon\ by $e^{-4K_1}$,
the next four by $1$ and the two last ones by $e^{4K_1}$, where
$K_1=-J_1/kT$.
To complete the definition of the model, we need to introduce a suitable 
order parameter for the folding transition. 
This parameter should distinguish between flat and
folded phases. Following \DGT, we introduce a fictitious magnetic
field $H$, coupled to the normal vectors to the faces, defined in
such a way that they all point up in a completely flat 
configuration of the lattice. The corresponding magnetization is
the order parameter for the folding transition. Setting $h=-H/kT$, 
the two last face configurations of Fig.\folon\ receive extra
Boltzmann weights $e^{\pm 4h}$ respectively, as one of these two flat 
faces points up and the other points down.
This model is studied in Sect.3 below.

\fig{The 32 possible face configurations of the rigid long edge model.
The variables $a,b$ may take any values mod 4, with $a\neq b$ mod 2.
This gives $8$ configurations for each of the first cases, and $4$
for each of the second and third ones. We have indicated the Boltzmann 
weights including a bending energy $K_2$ and a 
magnetic field $h$.}{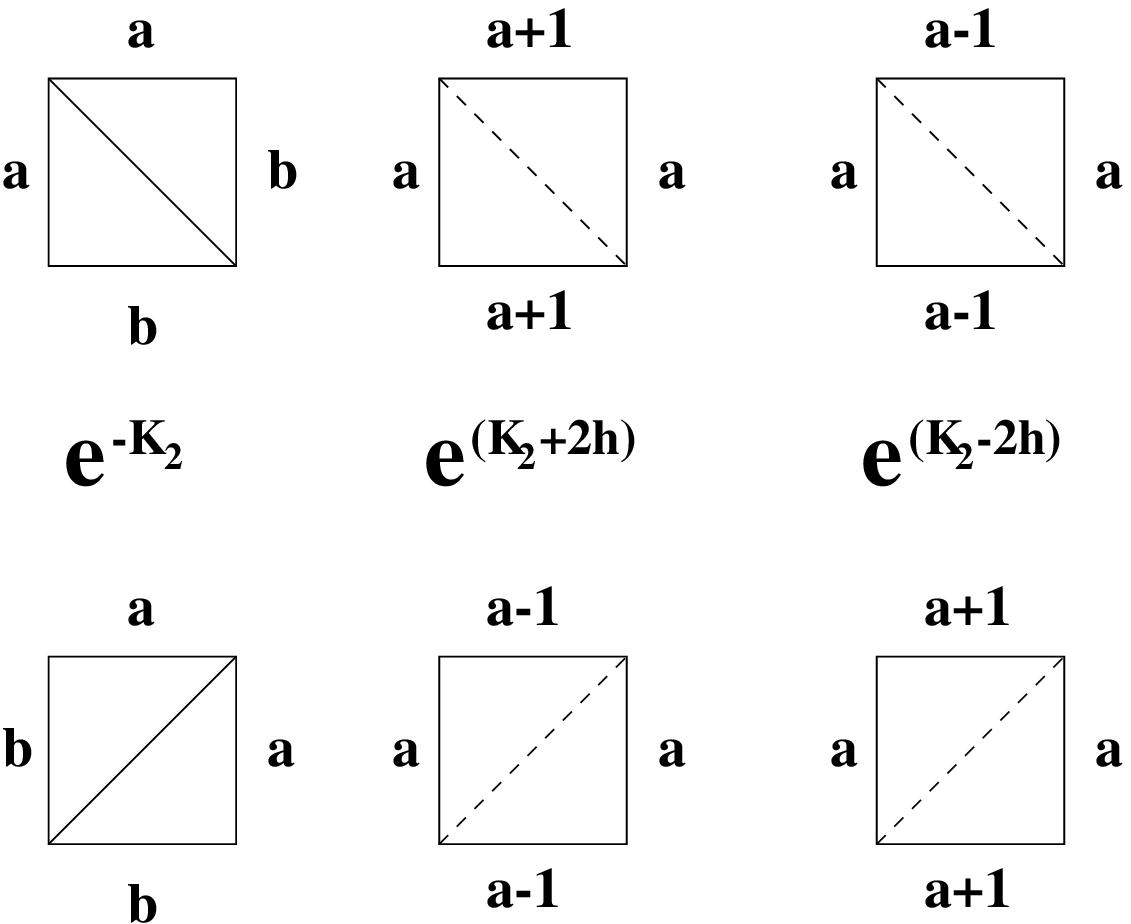}{7.cm}
\figlabel\folshor

On short edges, the constraint \consrho\
implies first that the two short edges of each triangle must
have perpendicular images. Using again a $\IZ_4$ short edge variable
reading respectively $0,1,2,3$ for the edge values 
$\vec{e_1},\vec{e_2},-\vec{e_1},-\vec{e_2}$, this means that
$a\neq b$ mod 2 for the two short edge values $a$, $b$ of 
a given triangle.
The constraint  also
relates the short edge images of any two triangles sharing a long edge
as follows. If $\vec{u}$, $\vec{v}$ stand for the short edge images
of the first triangle, and $\vec{u}'$, $\vec{v}'$ the second, there
are only two ways to accommodate $\vec{u}+\vec{v}=\vec{u}'+\vec{v}'$,
namely $\vec{u}'=\vec{u}$ and $\vec{v}'=\vec{v}$, or $\vec{u}'=\vec{v}$
and $\vec{v}'=\vec{u}$. 
The corresponding configurations are depicted in Fig.\folshor, for the
two types of faces (according to the position of the long edge).
In the first cases, the common long edge  
is folded (solid line), whereas in the second and third 
it is not (dashed line). Introducing a bending
energy $J_2$ to distinguish between folded and unfolded long edges, we
decide to weight the first cases with a Boltzmann weight $e^{-K_2}$,
and the second and third ones with a weight $e^{K_2}$, with $K_2=-J_2/kT$.
As before, we introduce a magnetic field $H$ coupled to the normal
vectors to the membrane. This results in an extra Boltzmann weight
$e^{\pm 2h}$ for respectively the second and third cases of Fig.\folshor,
with a flat face respectively pointing up and down.
This model is studied in Sect.4 below.

The general model including bending rigidity for both types of edges
is studied in Sect.6. In all cases, we will construct a row-to-row
transfer matrix $T$ for rows of width $L=1,2,...$ and extract its
largest eigenvalue, which dominates the partition function
\eqn\pfun{ Z~=~{\rm Tr}(T^M)~\sim~ \lambda_{max}^M~\sim~
e^{LMf}}
when $L$ and $M$ become large. This will lead to numerical estimates
for the thermodynamic free energy $f$ per site (note the change of 
sign and the absence of the usual $1/kT$ in the definition of $f$, 
for notational simplicity).
All the transfer matrices we are dealing with are sparse due to the
relatively short number of allowed face configurations, compared to
the size of the matrices. We therefore use an efficient algorithm
which codes the matrices as the list of their non-vanishing elements,
together with their position, and then extracts the largest eigenvalue
by repeated iteration on a given vector.

\newsec{Rigid Short Edge Model}

Including the abovementioned magnetic field $h=-H/kT$, the Boltzmann
weights for the face configurations of Fig.\folon\ read
\eqn\bowei{w_1=e^{-4K_1} \qquad w_2=w_3=w_4=w_5=1 \qquad  w_6=e^{4K_1+4h},\ 
w_7=e^{4K_1-4h} }
Note that the effect of the magnetic field is to distinguish the 
configurations in which the normal vectors all point up 
(energy $-4H$) or all point down (energy $4H$), from all other cases,
where there are as many up and down-pointing normal vectors.

For definiteness, let us denote
by $w(a,b,c,d)$ the Boltzmann weight attached to a cyclic configuration
$(a,b,c,d)$ of edge variables, clockwise around the square face
with $a$ in the west position.
The row-to-row transfer matrix of the model for a row of $L$ square faces
then reads
\eqn\transmaone{ T_{a_1,...,a_L;a_1',...,a_L'}~=~
\sum_{b_1,...,b_{L+1}\in \IZ_4}\prod_{i=1}^L w(b_i,a_i',b_{i+1},a_i) }
This matrix has size $4^L\times 4^L$ and is sparse, as there are
only $\sim 7^L$ non-vanishing elements (given the west value, there
are exactly 7 allowed face configurations). The numerical result 
of \DGT\ for the entropy of folding $f=s_{SD}\simeq .230...$ is 
readily recovered for $K_1=h=0$.

\fig{Magnetization versus magnetic field for the rigid short edge
model at $K_1=0$, for strips of width $L=1,2,...,6$. We see clearly
two phases: folded, with $M\to 0$, and flat, with $M=1$.
We expect the transition between the two to be discontinuous in the 
thermodynamic limit $L\to \infty$.}{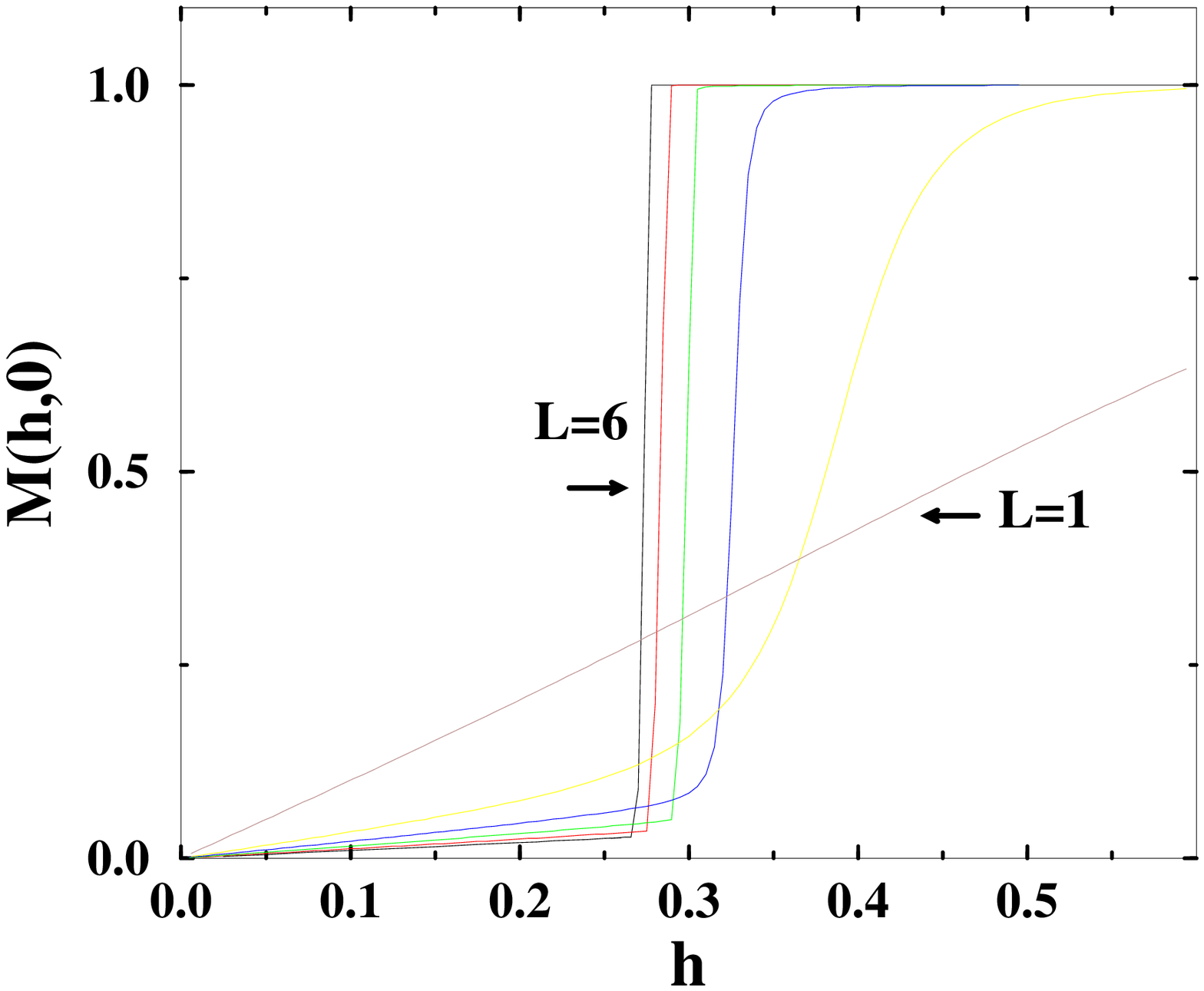}{7.5cm}
\figlabel\magone

We display in Fig.\magone\ the magnetization $M=\partial_h f$ for
various sizes $L=1,2,...,6$ at $K_1=0$. 
Even for these finite values of $L$, we see a drastic jump in
the magnetization at a value $h=h_{c,L}(K_1=0)$ of the magnetic field.
Extrapolating these, we find that they tend to a limiting value 
$h_c(0)=.23...$ for large $L$. In the low-$h$ phase, the 
magnetization has a small finite slope, which tends to $0$ for 
large $L$. In the large-$h$ phase, the magnetization is identically $1$.
We conclude that the system is reduced to two phases,
a completely folded one for $h<h_c(0)$ and a completely flat one
for $h>h_c(0)$. Note that $h_c(0)\simeq s_{SD}$.

\fig{The magnetization as a function of the magnetic field
for strips of width $L=6$, and various values of $K_1=.5,.25,.1,0,-.1$.
A change of behavior is observed for $K_1=.5$: there,
we expect the folded phase to 
disappear in the thermodynamic limit.}{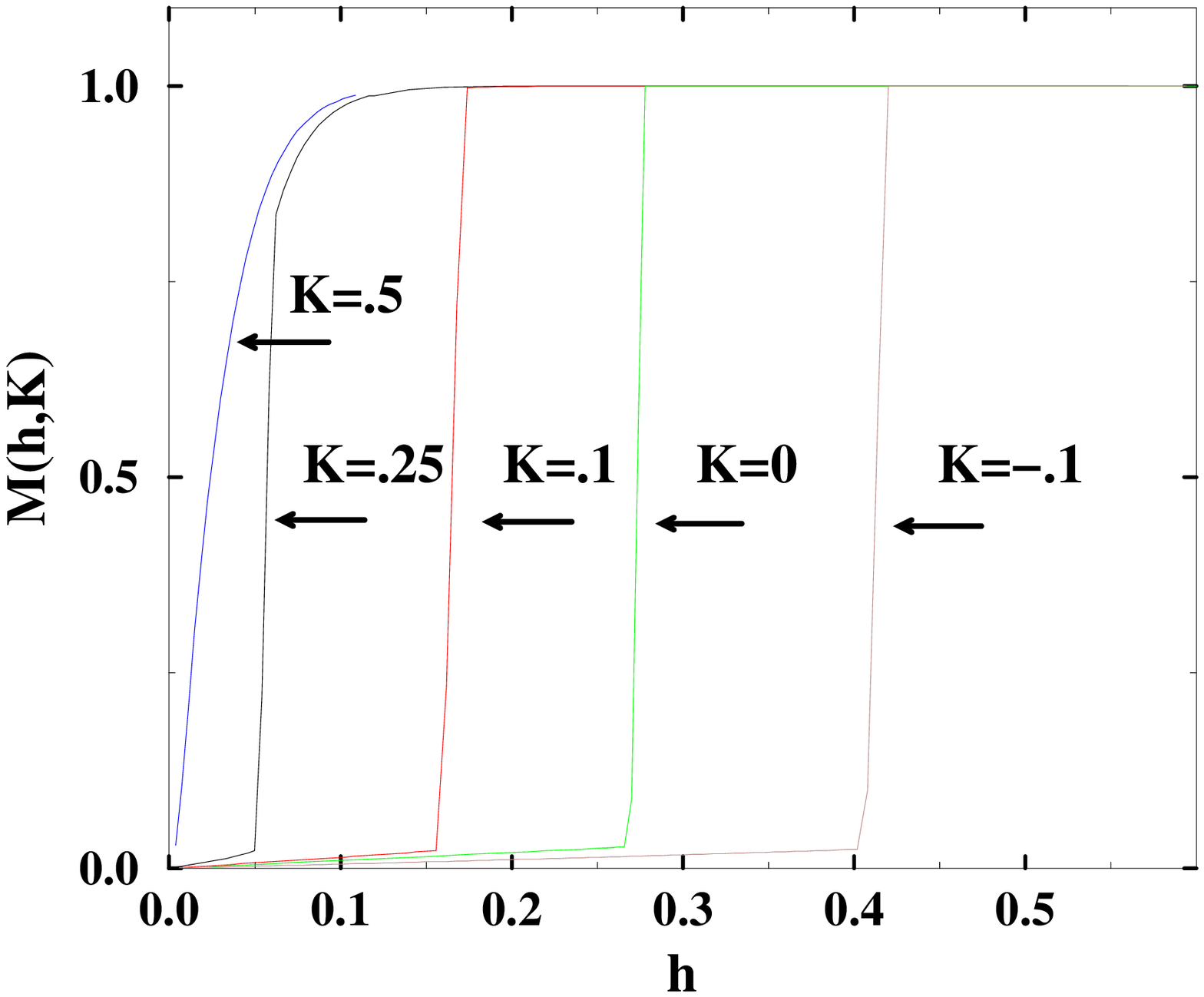}{7.5cm}
\figlabel\momagone

This two-phase situation is observed for other values of $K_1$ in
Fig.\momagone, where we have represented the corresponding
magnetizations as functions of $h$ for
$L=6$. Like in \DGT, this leads us to 
formulate the hypothesis that, at least for small enough $K_1$,
there will always exist only two phases, folded and flat, with $M=0$ and $1$
respectively. 
Let $f_0(K_1)$ denote the free energy per triangle of
the model in zero magnetic field.  This is exactly the energy of the
folded phase of our model, as the latter is insensitive to the magnetic
field. Note that this phase is entropic, at least when $K_1=0$, 
where $f_0(0)=s_{SD}=.230...$ 
On the other hand, the completely flat phase consists
of configurations made only of faces of the form $(a,a+1,a+2,a-1)$,
with weight $w_4=e^{4(K_1+h)}$, hence resulting in a thermodynamic
energy $f_1(K_1,h)=K_1+h$ per triangle. 
The critical field $h_c(K_1)$
is then defined by the equality of the two energies, namely
\eqn\crih{ h_c(K_1)~=~f_0(K_1)-K_1 }

\fig{The critical field $h_{c,L}(K)$ for L=1,2,...,7 (from top to bottom).
We have represented in dashed line the extrapolated limiting curve
for $L\to \infty$. It displays a transition  
at $K=K_{c1}\simeq .27$.}{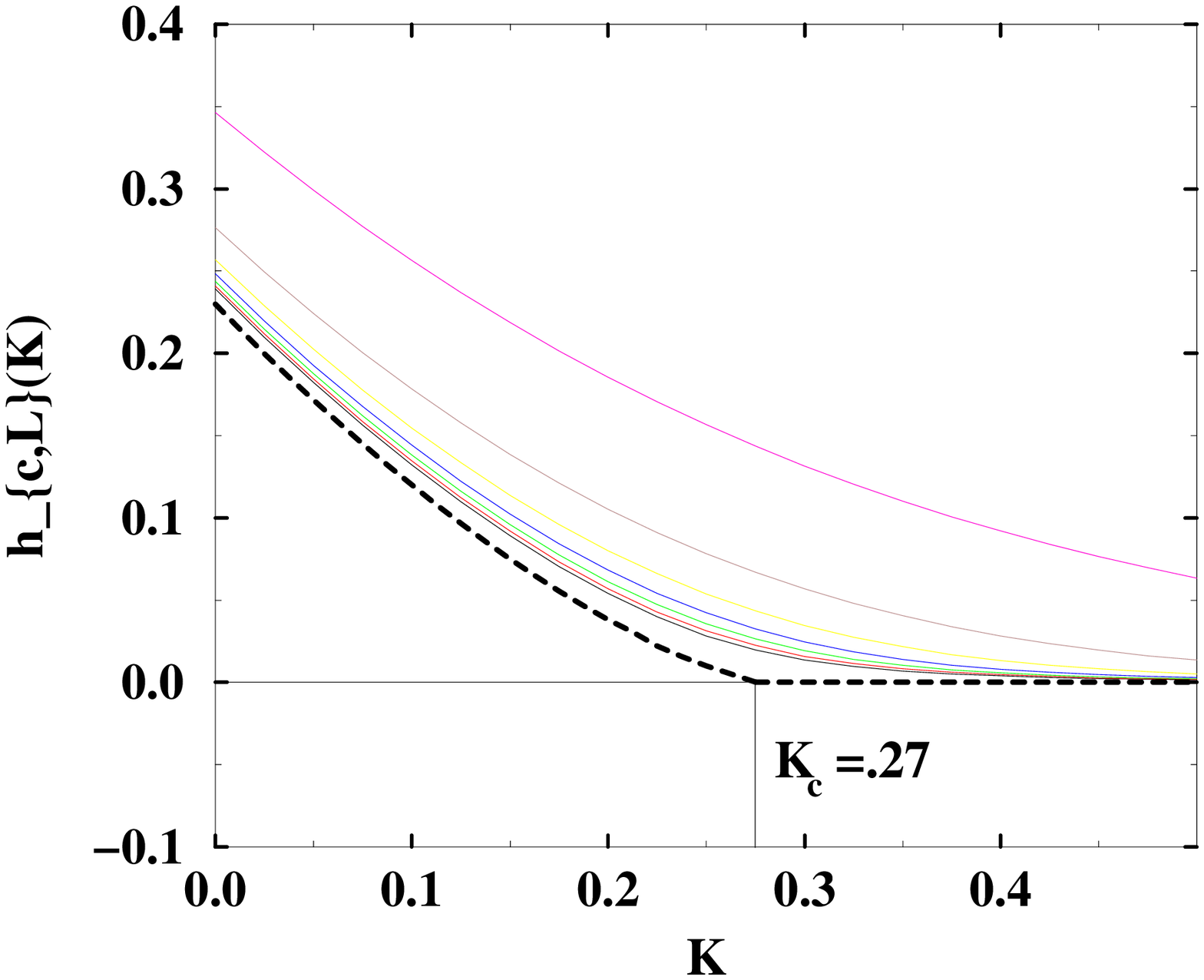}{8.cm}
\figlabel\hcrone

Let $\lambda_{max}^{(L)}(K_1)$ denote the largest (Perron-Frobenius)
eigenvalue of the transfer matrix $T$ of \transmaone.
Taking $\phi_{0,L}(K_1)~=~{\rm Log}(\lambda_{max}^{(L)}(K_1))/(4L)$ as
a sequence of approximations for $f_0(K_1)$, the folded phase energy,
is not very accurate, as it includes a generic two-fold degeneracy 
due to the invariance under reversal of all non-horizontal
tangent vectors, i.e. a folding along the top or bottom horizontal
line of the row (in which, in particular,
the two face configurations $(a+1,a+2,a-1,a)$ and $(a-1,a+2,a+1,a)$
are exchanged).
We should rather use
\eqn\bestap{ f_{0,L}(K_1)~=~\phi_{0,L}(K_1)-{1\over 4L}{\rm Log\, 2} }
by explicitly dividing out the degeneracy.
Although in principle equivalent to $\phi_{0,L}(K_1)$ in the thermodynamic
limit, $f_{0,L}$ displays a much better convergence.
Let us now turn to the finite $L$ flat phase energy 
$f_{1,L}(K_1,h)$. 
The flat phase matrix elements of $T$ are non-diagonal, and read
$T_{a,a+2,a,a+2,...;a+2,a,a+2,a,...}=2e^{4LK_1}\cosh(4Lh)$. They
give the energy
\eqn\flaen{ f_{1,L}(K_1,h)~=~K_1+h+{1\over 4L}{\rm Log}(1+e^{-8Lh})}
Due to the exponential character of the correction,  
the approximation to the critical field, $h_{c,L}(K_1)=f_{0,L}(K_1)-K_1$
is accurate, at least in the large $h$ phase. 
We have represented
in Fig.\hcrone\ the critical field $h_{c,L}(K)$ as a function of $K$,
for various values of $L$. 
These curves are extrapolated to the large $L$ limiting curve 
represented in dashed line. This curve has also been obtained by the
extrapolation of another sequence of approximations, namely
\eqn\otzseq{ h_{c,L/L+1}(K)={1\over 4}{\rm Log}\bigg({\lambda_{max}^{(L+1)}(K)
\over \lambda_{max}^{(L)}(K)}\bigg) -K }
for $L=1,2,...,6$, with an excellent agreement.
We observe the existence of a finite value
$K_{c1}=.27$ of $K_1$ beyond which $h_c$ vanishes identically.

\newsec{Rigid Long Edge Model}

Including the magnetic field $h=-H/kT$, the Boltzmann weights of
the configurations of Fig.\folshor\ read respectively
\eqn\bolweto{ w_1=e^{-K_2} \qquad w_2=e^{K_2+2h} \qquad w_3=e^{K_2-2h}}
As before, let $w(a,b,c,d)$ stand
for the Boltzmann weight of the configuration $(a,b,c,d)$ around
the face, clockwise with $a$ in the western position.  
By a slight abuse of notation, we  will use the same letters to denote
the Boltzmann weights, transfer matrix, energy, etc... of the present
model as those used for the Rigid Short Edge model of previous section.

With these definitions, the transfer matrix for a row of width $L$
reads
\eqn\tmato{ T_{a_1,...,a_L;a_1',...,a_L'}(h,K)~=~
\sum_{b_i,c_i,d_i\in \IZ_4} \prod_{i=1}^L 
w(b_i,c_i,b_{i+1},a_i)w(d_i,a_i',d_{i+1},c_i) }
Indeed, because of the alternance of the long edge between first and
second diagonal in the short square faces, we must consider two rows of 
square faces, with a total of $4L$ triangles. Let us note however that
$T=U(h,K) U^t(-h,K)$, where $U(h,K)$ is the transfer matrix for the 
first row, with entries
\eqn\firo{ U_{a_1,...,a_L;a_1',...,a_L'}(h,K)~=~
\sum_{b_i\in \IZ_4} \prod_{i=1}^L 
w(b_i,a_i',b_{i+1},a_i) }
Indeed, upon transposition, the picture for $U$ is reflected wrt a
horizontal line, hence the inner long edge on each face switches to the
other diagonal, which amounts to a change $h\to -h$ 
according to Fig.\folshor.

%table I
\par\begingroup\parindent=0pt
\leftskip=1cm\rightskip=1cm\parindent =0pt
\baselineskip=11pt
$$\vbox{\font\bidon=cmr8 \def\bidondon{\bidon} \bidondon \offinterlineskip
\halign{\tv \quad # \tv
& \hfill \ #
& \hfill # \tv \cr
\noalign{\hrule}
\tvi  $n$ \hfill & $\lambda_{max}$ \hfill & $\nu_n$ \hfill\cr
\noalign{\hrule}
\tvi 1 & 4.000000 & 1.22891 \cr
\tvi 2 & 6.040895 &  1.23758 \cr
\tvi 3 &  9.252396 & 1.24610 \cr
\tvi 4 &  14.36690 & 1.24926 \cr
\tvi 5 & 22.42177 & 1.25212 \cr
\tvi 6 & 35.15328 &  1.25351 \cr
\tvi 7 & 55.23640 & 1.25482 \cr
\tvi 8 & 86.97476 & 1.25558 \cr
\tvi 9 & 137.1155 &  1.25631 \cr
\tvi 10 & 216.4118 &   \cr
\noalign{\hrule}
}}$$
{\bf Table I:} Numerical results for the maximum eigenvalue
of the square root of the transfer matrix
of the rigid long edge model with free boundary conditions. 
We have represented the size $n$ of the row, the largest eigenvalue
and the ratio $\nu_n=(\lambda_{n+1}/\lambda_n)^{1/2}$, which
converges to the partition function per triangle.
\par
\endgroup\par

This remark makes the numerical study much simpler, as the extraction of
the largest eigenvalue of $T$ is done by iterating the action of $U$
and $U^t$ in alternance on a given vector. Again, $U$ is sparse, with 
a number of non-zero elements $\sim 2^L$ much smaller that its size $4^L$. 
As an application, we list in Table I the largest eigenvalue of the
matrix $\sqrt{T}=\sqrt{UU^t}$ for $K_2=h=0$, and for $L=1,2,...,10$.
This is in agreement with the estimate for the partition function 
per triangle $z_{SD}=1.258...$ of \PDF.

\fig{Magnetization versus magnetic field for strips
of width $L=7$ in the rigid long edge model, for various values
of $K$. We note the existence of two phases with respectively
$M=0$ (folded) and $M=1$ (flat).}{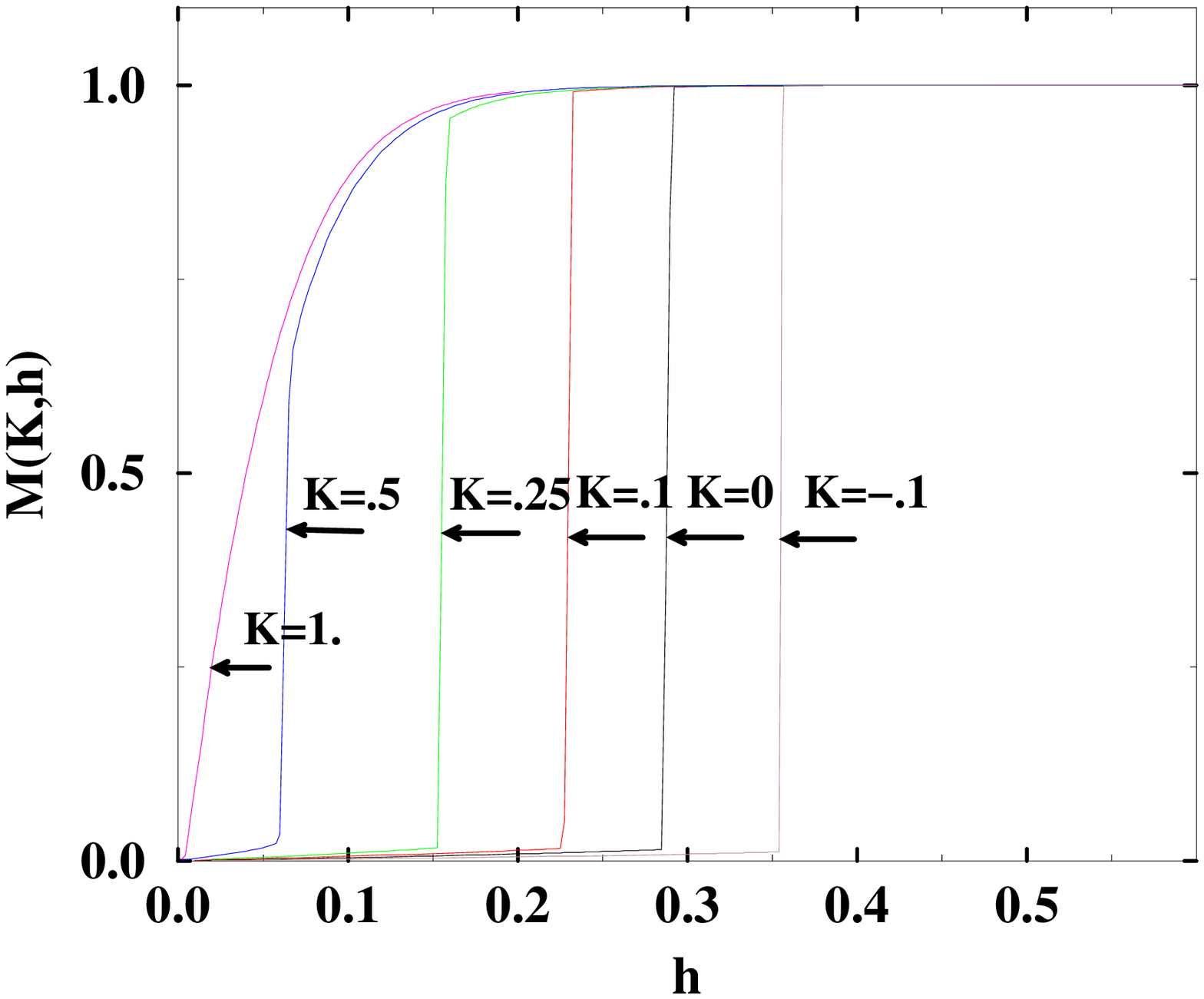}{8.cm}
\figlabel\confirto

The two-phase hypothesis of the previous section is confirmed here
in Fig.\confirto, where we have represented the magnetization as
a function of $h$ for $L=7$ and various values of $K_2$.
An analogous reasoning leads to the determination of
the critical magnetic field $h_c(K_2)=f_0(K_2)-{K_2\over 2}$, obtained
by equating the energies of the folded and flat phases. Here the
flat phase is made of only the face configuration $(a,a+1,a,a+1)$,
hence an energy of $f_1(K_2,h)=(2L(K_2+2h))/(4L)={K_2\over 2}+h$.

\fig{The critical fields $h_{c,L}(K)$ for $L=1,2,...,9$.
We have represented the extrapolated large $L$ limit in thick dashed
line. It displays a transition at the critical 
value $K_{c,2}\simeq .54$.}{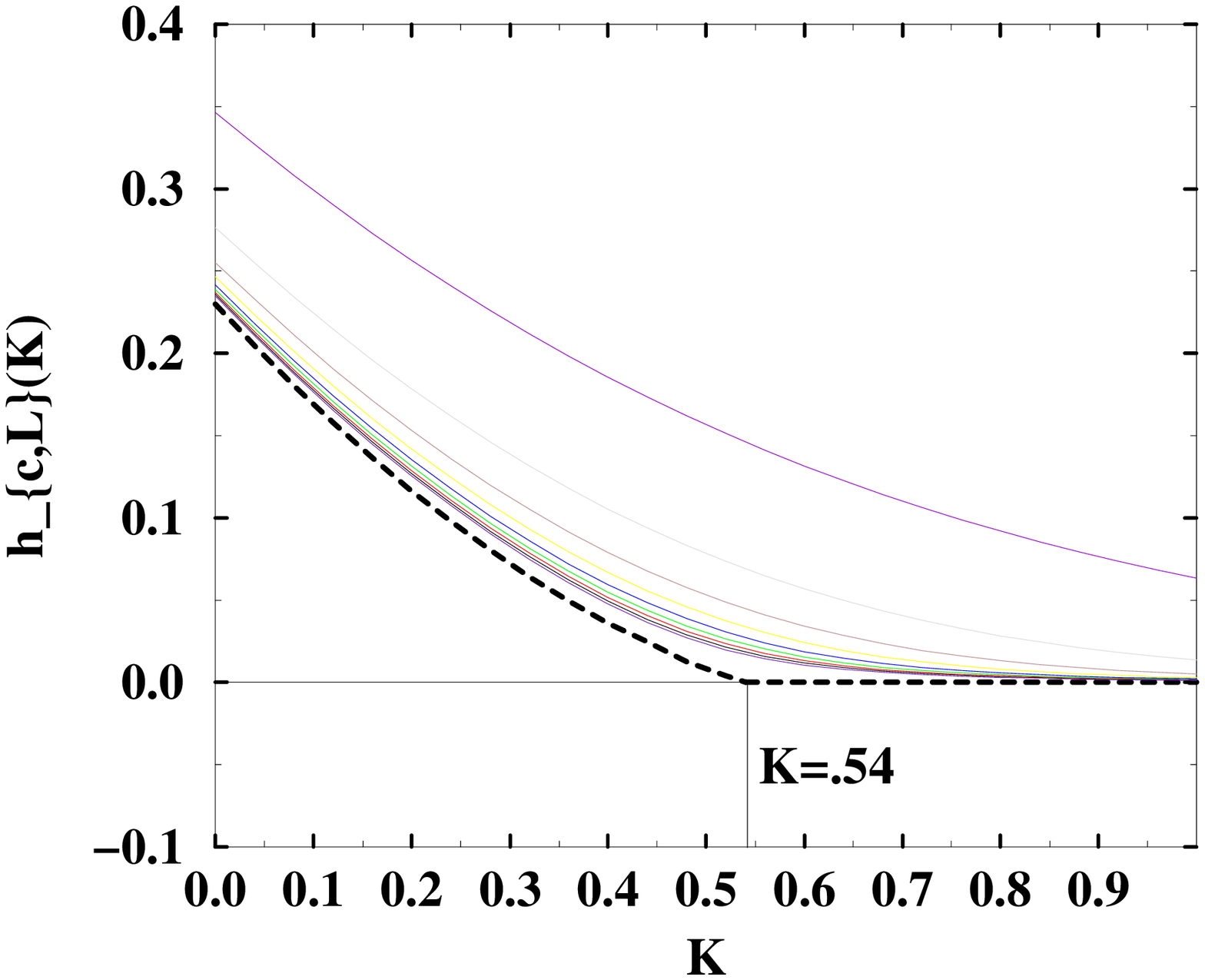}{8.cm}
\figlabel\hcrito

For strips of finite size $L$ however, the folded phase energy 
reads 
\eqn\frenL{f_{0,L}(K_2)~=~{1\over 4L}{\rm Log}\lambda_{max}^{(L)}(K_2)
-{1\over 4L}{\rm Log}\, 4} 
where $\lambda_{max}^{(L)}(K_2)$ is the largest (Perron-Frobenius)
eigenvalue of the transfer matrix $T$ \tmato\ for a row of 
$2\times L$ squares. Note the subtraction of the $4=2\times 2$ times 
degeneracy under the previously mentioned 
reversal of non-horizontal tangent vectors, in which in particular 
$(a+1,a,a+1,a)\leftrightarrow (a-1,a,a-1,a)$.
On the other hand, the flat phase energy 
corresponds to the
diagonal matrix elements $T_{a,a,a,...;a,a,a...}$, $a\in \IZ_4$,
with only face configurations of the type $(b,a,b,a)$ and $b\neq a$
mod 2.
This gives four identical diagonal blocks, with eigenvalue 
\eqn\maxev{ \lambda_{max}^{(L)}(K_2,h)~=~2 e^{2LK_2} \cosh(4Lh) }
and the energy
\eqn\flen{ f_{1,L}(K_2,h)~=~{K_2\over 2}+h+{1\over 4L}{\rm Log}(1+e^{-8Lh}) }
Neglecting the exponential correction,
the equality between \flen\ and \frenL\ yields the finite size
critical field
\eqn\finhcri{ h_{c,L}(K_2)~=~f_{0,L}(K_2)-{K_2\over 2} }
which we have represented for various values of $L$ in Fig.\hcrito.
We observe an excellent convergence of the quantity $h_{c,L}(K_2)$,
and a particular value $K_{c,2}\simeq .54$ of $K_2$ beyond which the critical 
field $h_c$ vanishes identically, hence only the flat phase survives.
Note that we get $K_{c,2}\simeq 2K_{c,1}$.

\newsec{Continuous Transitions for $K<0$}

As observed in \DGT\ in the case of the triangular lattice folding,
both short and long rigid edge models undergo continuous phase
transitions for some negative critical values $K_{*,1}$ and
$K_{*,2}$ of $K$. In terms
of our initial setting, these are anti-ferromagnetic transitions.
However, they take place in the entropic (folded) phase of both models,
in which local excitations can be formed, as explained in \PDF.

\fig{A local excitation of one of the anti-ferromagnetic (completely folded)
groundstates of the rigid short edge model. The unfolded
short edges are represented in dashed line. All others remain
folded.}{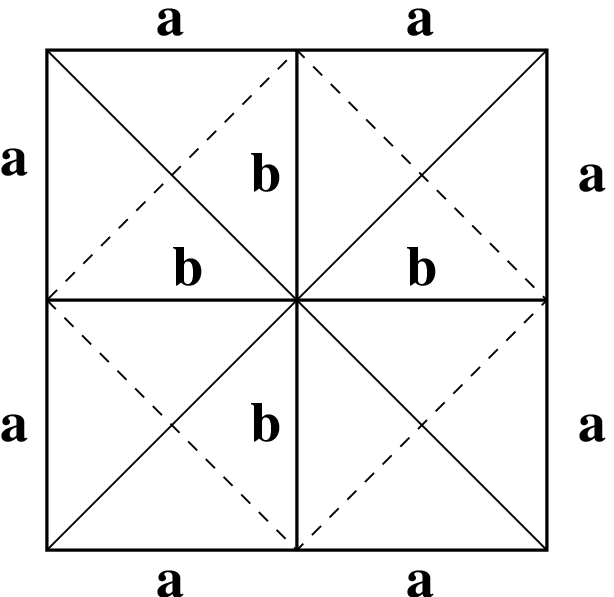}{5.cm}
\figlabel\unfosq

More precisely, in the rigid short edge model, starting from one 
of the (completely folded) anti-ferromagnetic groundstates, in which
all the short edges are completely folded, hence where 
all long edges take the same value $a$, we can unfold a square of
8 short edges, as shown in Fig.\unfosq. This is a local excitation 
of the anti-ferromagnetic groundstate, which receives a relative
Boltzmann weight $2 e^{-16K_1}$, including the effect
of unfolded edges and the free choice of the two orientations
$b=a\pm 1$ of the new tangent vectors.
Note that we have not specified the folding state of the long
edges, which results in a non-vanishing entropy of the model.
Such local excitations can be combined so as to form any loops
of unfolded short edges (with the constraint that these edges can 
only be unfolded by parallel pairs within each square of long edges).

\fig{A local excitation of one of the anti-ferromagnetic
(completely folded) groundstates of the rigid long edge model
(with $a\neq b$ mod 2).
The four dashed long edges have been unfolded in the process,
while all others remain folded.}{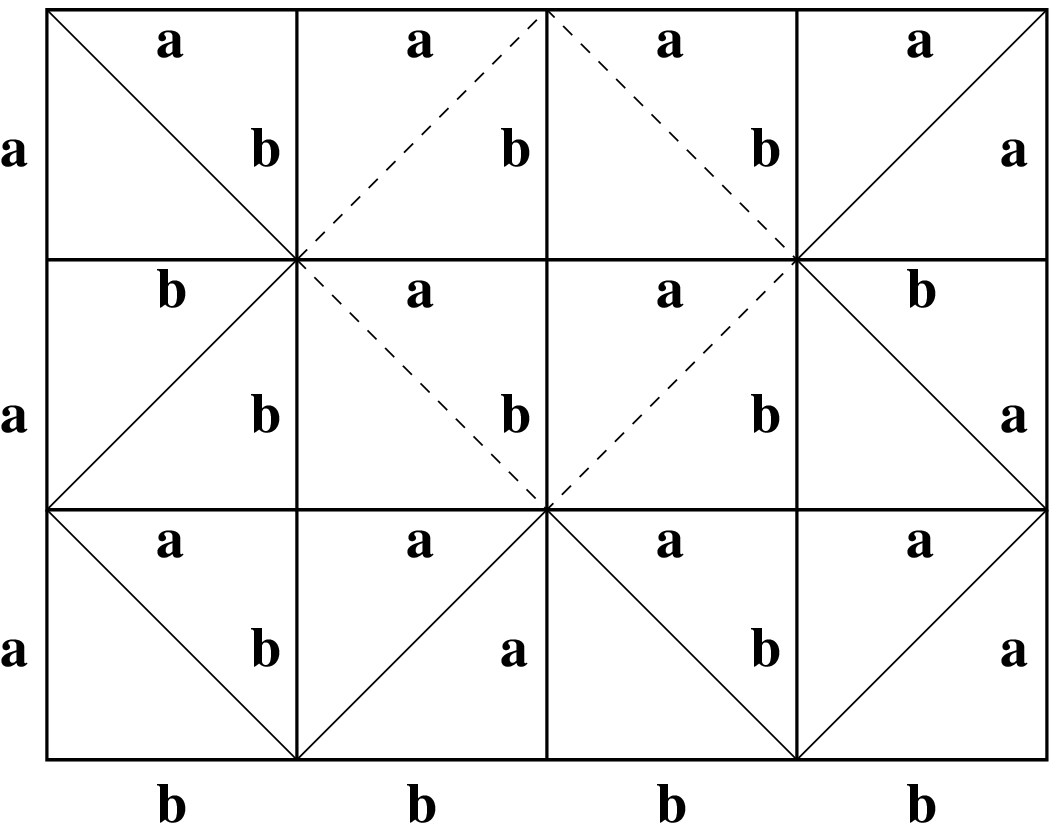}{7.cm}
\figlabel\unfolon

In the rigid long edge model, similar excitations of the (completely 
folded) anti-ferromagnetic groundstate may occur. They correspond 
to unfolding four long edges forming a square, as shown in 
Fig.\unfolon. This excitation receives a relative Boltzmann weight
$2e^{-8K_2}$. Again, these excitations may be combined 
into any loop of unfolded long edges. 

\fig{Specific heat vs K for the rigid short (1) and long (2) edge models.
The pairs $L/L+1$ used to evaluate the curves are indicated. Both
models display a continuous transition. The critical values
of $K$ are estimated to be $K_{*,1}=-.08$ and $K_{*,2}=-.1$.
}{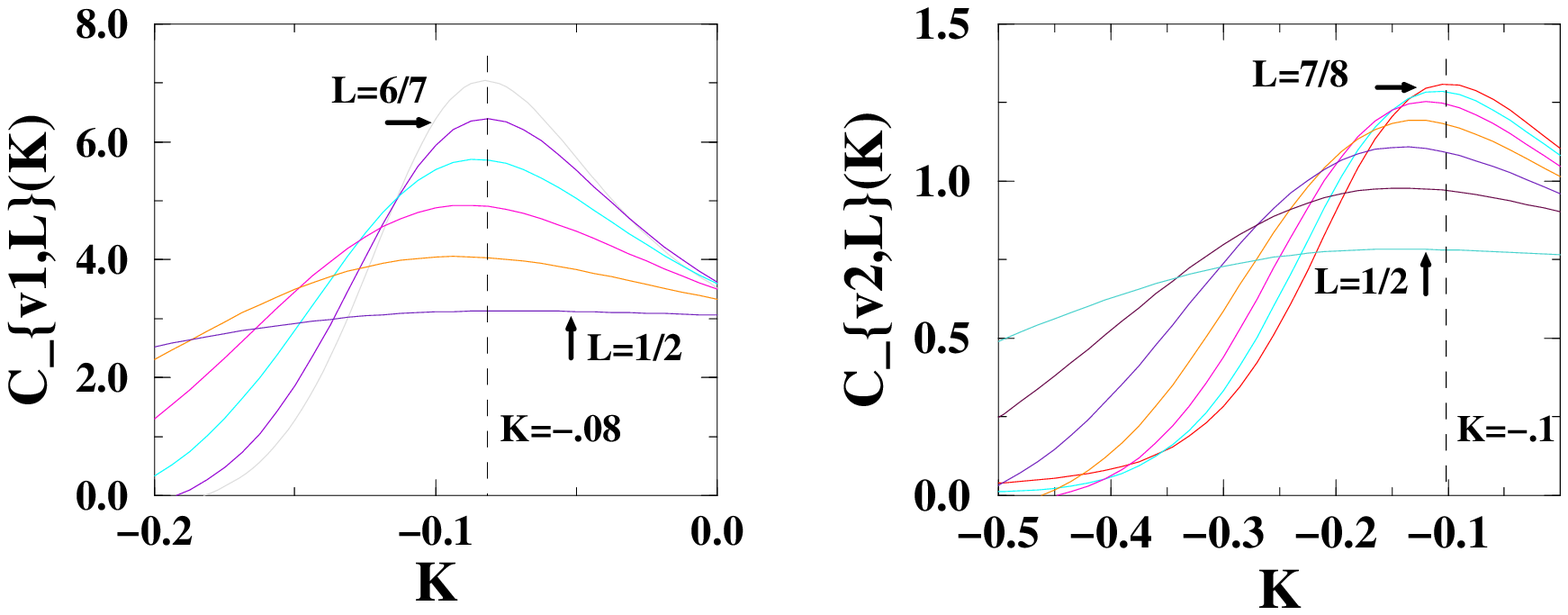}{12.cm}
\figlabel\cvs

In both cases, the resulting loop model is naturally expected to undergo
a phase transition of the Ising type, when $K_1,K_2<0$ cross some
critical values. To check this, we have represented 
in Fig.\cvs\ for $L=1,2,...,6$ or $7$ the specific
heats $C_{v1,L}(K_1)=\partial_{K_1}^2\varphi_{1,L}(K_1)$ and 
$C_{v2,L}(K_2)=\partial_{K_2}^2\varphi_{2,L}(K_2)$, where the
approximations to the free energy are taken to be
\eqn\aprofre{ \varphi_{i,L}(K_i)~=~ {1\over 4}{\rm Log}\bigg(
{\lambda_{max}^{(L+1)}(K_i)\over \lambda_{max}^{(L)}(K_i)}\bigg) }

We observe the typical growth of the maximum of the specific
heat with $L$, which points to critical values of $K$ reading
respectively $K_{*,1}=-.08$ and $K_{*,2}=-.1$.
Using the graphs of Fig.\cvs,
we have also obtained estimates for the thermal critical exponents $\alpha_i$
such that in the thermodynamic limit
$C_{vi}(K_i)\simeq |K_i-K_{*,i}|^{-\alpha_i}$, with the results
$\alpha_1\simeq .2$ and $\alpha_2\simeq .25$.

\newsec{General Phase Diagram}

Let us now address the general rigid (long and short) edge model
for the square-diagonal lattice folding problem.
This is simply a combination of the two models of Sects.3 and 4,
where the long and short edges receive the Boltzmann weights 
$e^{\pm K_1}$ (short) and $e^{\pm K_2}$ (long) for unfolded ($+$) 
and folded ($-$) states of the edge.

In view of the results of Sects.3,4 and 5, we expect a very interesting 
phase diagram for this general model, in which both first order and 
continuous transition can be found.

\fig{The row-to-row transfer matrix $T$ for the general
rigid long and short edge model. $T$ acts by transferring a 
zig-zag row $\{(a_1,b_1),(a_2,b_2),...,(a_L,b_L)\}$ of pairs of
adjacent short edges to the nearest vertical translate.}{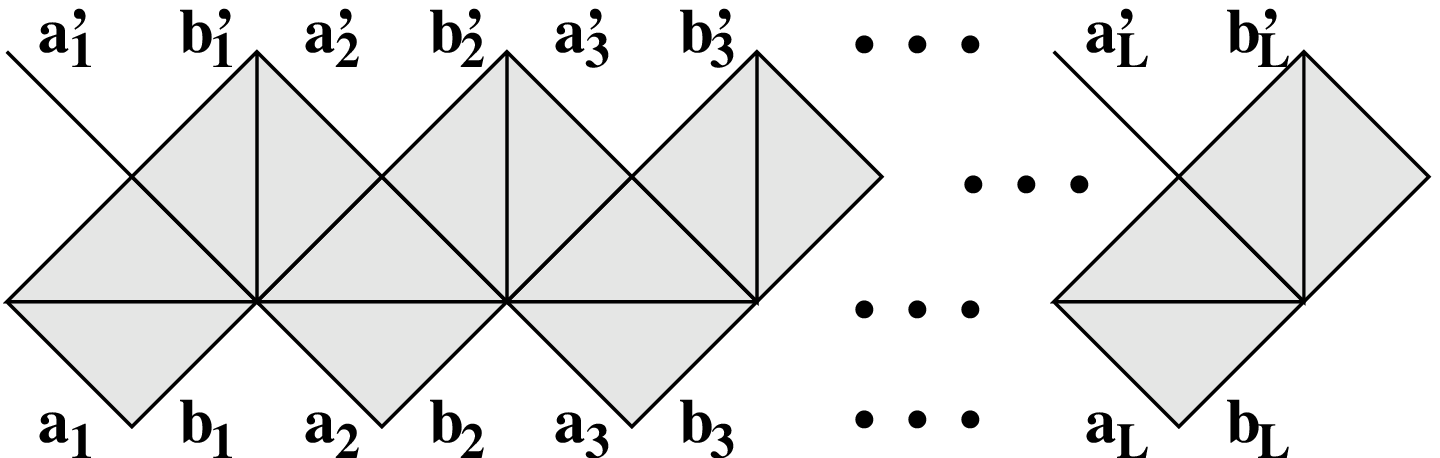}{8.cm}
\figlabel\row

The transfer matrix $T$ of the general model reads as follows. We consider a row
of $4L$ triangles, as shown in Fig.\row, which transfers a zig-zag line
of $2L$ short edges to the nearest translate above it. The short edges
in a line naturally go by pairs of adjacent edges sharing the same triangle,
with a total of $8$ possible images, $a,b\in \IZ_4$, with $a\neq b$ mod 2,
expressing the fact that the corresponding tangent vectors are orthogonal. 
Hence we may view $T$ as a $8^L\times 8^L$ matrix acting in the space of pairs
of adjacent short edges along a zig-zag line of $2L$ edges.

\fig{The transfer matrix of the rigid long and short edge model
is made of three elementary Boltzmann weights $w_1,w_2,w_3$ corresponding
to the arrangements of inner short edges $c,d,e\in \IZ_4$, with
$c\neq d$ mod 2, and outer short edges $a,b,a',b',a''\in \IZ_4$,
with $a\neq b$ mod 2, $a'\neq b'$ mod 2.}{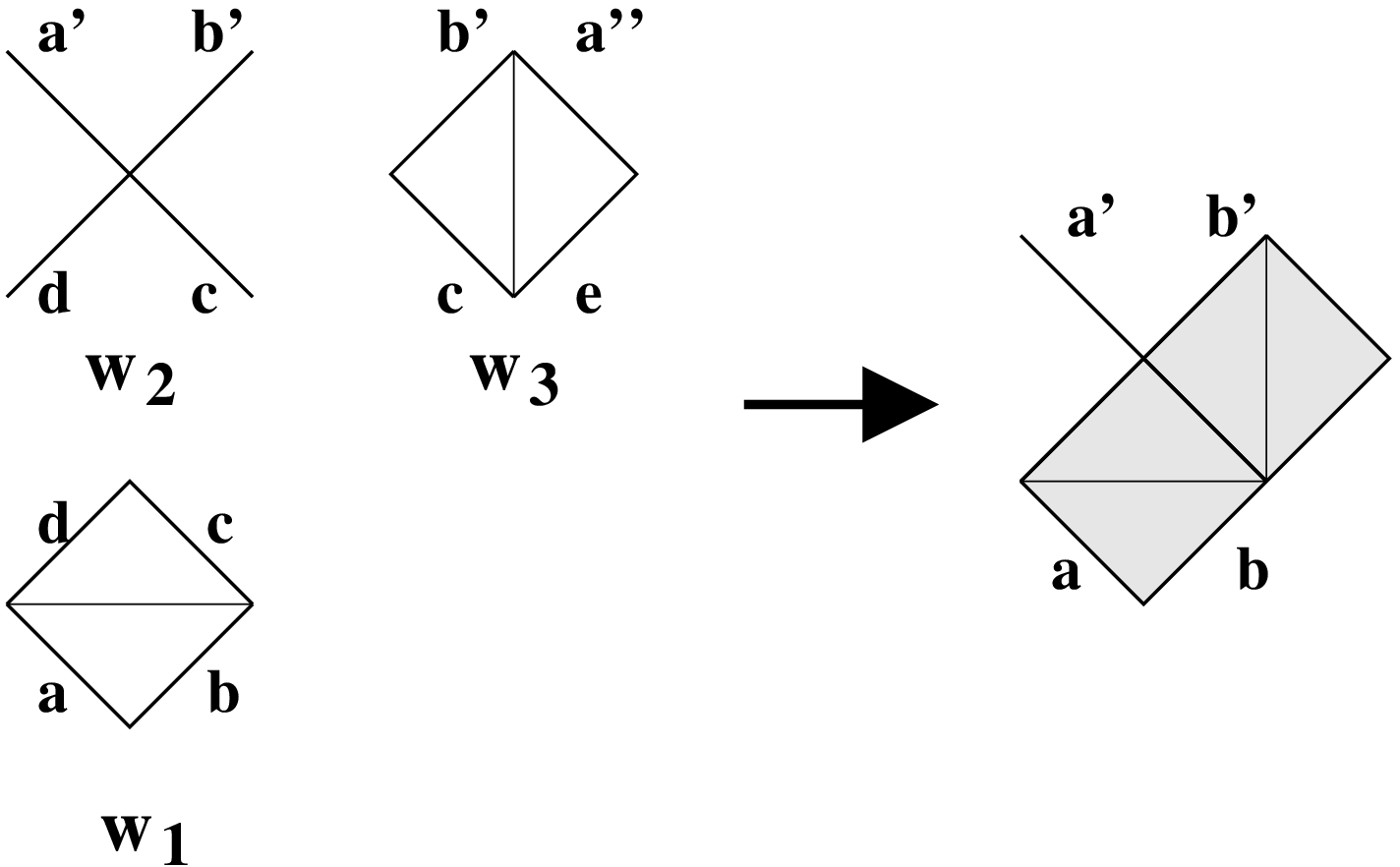}{7.5cm}
\figlabel\bit

The matrix $T$ is made of $L$ cells, expressing the various Boltzmann weights
and constraints on the short and long inner edges. We distinguish three pieces,
according to Fig.\bit,  with the Boltzmann weights
\eqn\introw{\eqalign{
w_1(a,b,c,d)~&=~ e^{K_2} \delta_{a,c}\delta_{b,d} 
+e^{-K_2} \delta_{a,d}\delta_{b,c}\cr
w_2(a',b',c,d)~&=~ (e^{2K_1}\delta_{a',c+2}+e^{-2K_1}\delta_{a',c})
(e^{2K_1}\delta_{b',d+2}+e^{-2K_1} \delta_{b',d}) \cr
w_3(b',a'',e,c)~&=~e^{K_2} \delta_{b',e}\delta_{a'',c} 
+e^{-K_2} \delta_{a'',b'}\delta_{e,c}\cr}}
where $\delta_{p,q}=1$ if $p=q$ mod 4, $0$ otherwise.
The transfer matrix elements then read
\eqn\transelg{\eqalign{
T_{\{ a_i,b_i\};\{ a_i',b_i'\} }~&=~ \cr
\sum_{c_i,d_i, a_{L+1}',d_{L+1}\in \IZ_4 \atop 
i=1,2...L} \prod_{i=1}^L
&w_1(a_i,b_i,c_i,d_i) w_2(a_i',b_i',c_i,d_i) 
w_3(b_i',a_{i+1}',d_{i+1},c_i) \cr}}

Inspecting \introw, we see that $T$ has $\sim 16^L$ 
non-vanishing entries,
for a size $8^L\times 8^L$, hence is a sparse matrix. But the size is of course
much more limiting than in the cases of Sects.3 and 4.

\fig{Maxima of the specific heat $C_{v,L}(K_1,K_2)$ for $L=2$ (stars),
$L=3$ (circles) and $L=4$ (diamonds), in the $(K_1,K_2)$ plane.
We have represented the axes as well as the line $K_2=-2K_1$ in
dashed lines. The latter appears to be the separation between the continuous
(anti-ferromagnetic) transition curve and the first order 
(ferromagnetic) one.}{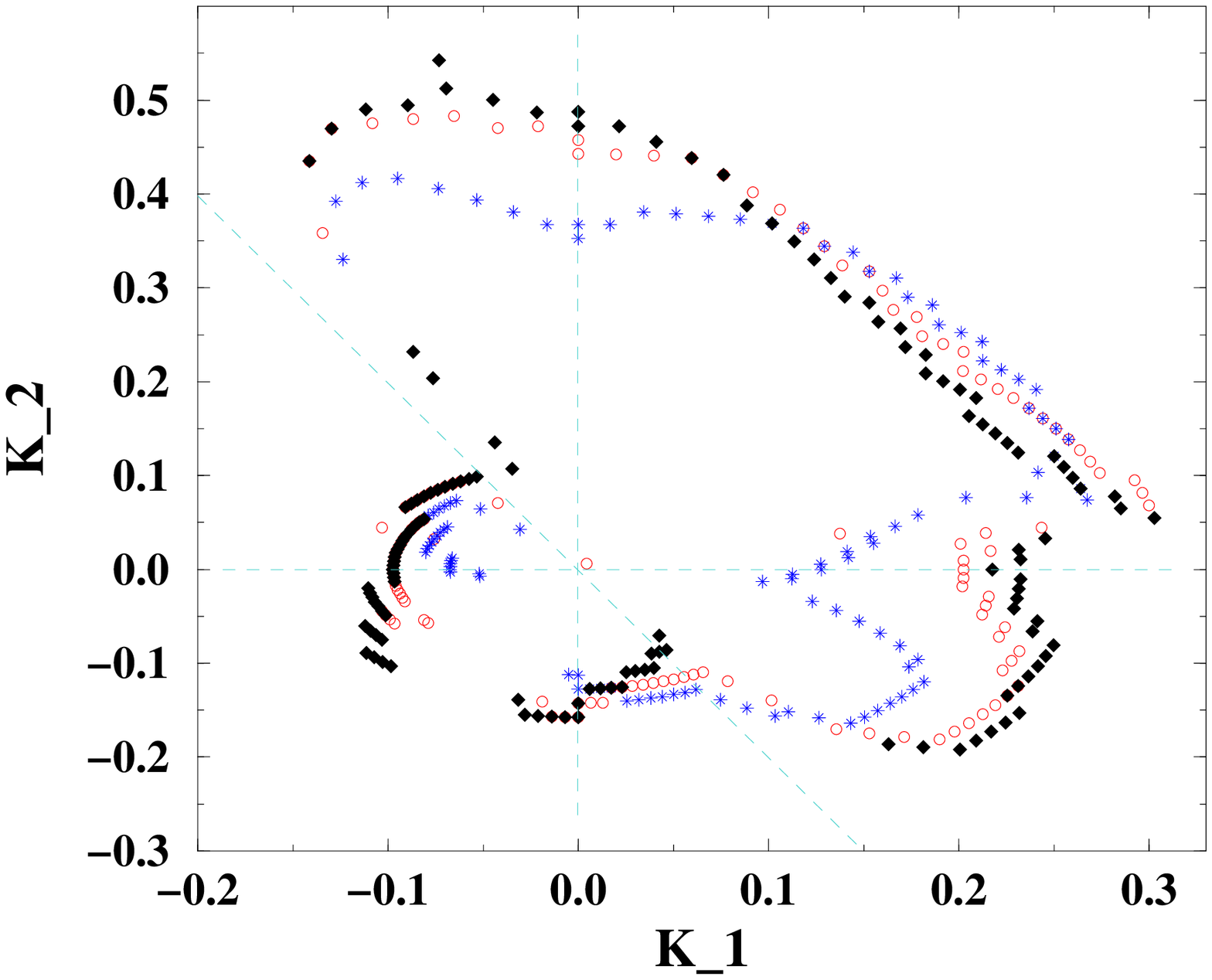}{9.cm}
\figlabel\maxcv

The transition curves we are looking for
are characterized numerically, for finite size $L$ 
of the transfer matrix, by a growing peak of the specific heat.
To identify those in the $K_1,K_2$ plane, we have located the
maxima of the specific heat along lines through the origin. More precisely,
taking $K_1=K\cos(\theta)$ and $K_2=K\sin(\theta)$, we have taken a discrete
set of angles $\theta\in [-\pi/2,\pi/2)$, and represented the maxima of
the specific heat
\eqn\spehl{ C_{v,L}(K)~=~\partial_K^2 {1\over 4L}{\rm Log}\bigg(
\lambda_{max}^{(L)}(K\cos(\theta),K\sin(\theta))\bigg)}
where $\lambda_{max}^{(L)}(K_1,K_2)$ denotes the maximum (Perron-Frobenius)
eigenvalue of the transfer matrix $T$ defined above, for a strip of width $L$.
The results for $L=2,3,4$ and $\theta=k\pi/(2m)$, $k=-m,-m+1,...,m-2,m-1$,
$m=35$, are displayed in Fig.\maxcv.

\fig{The specific heat $C_{v,4}(K_1,-2K_1)$ for $L=4$, along
the crossing line $K_2=-2K_1$. We have indicated the transition points
(or remnants thereof).}{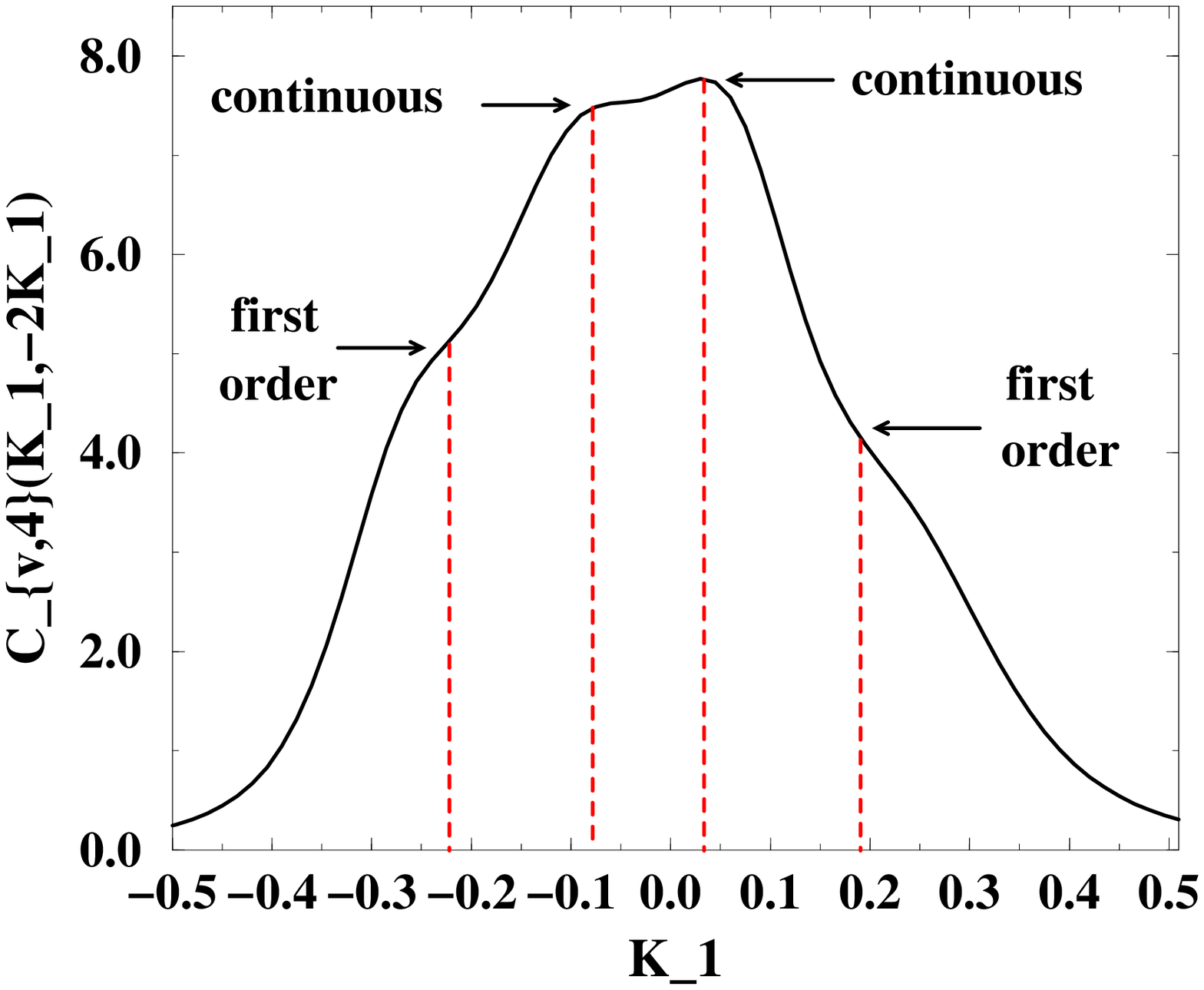}{7.cm}
\figlabel\koneto

Note that the separation between the two transition curves,
continuous anti-ferromagnetic (negative $K$'s) and first order ferromagnetic
(positive $K$'s) appears to lie precisely on the line
$K_2=-2K_1$. This should not be too much of a surprise, if we note that
there are exactly twice as many short edges as long ones in the square-diagonal
lattice. The anti-ferromagnetic and ferromagnetic bending
energies are  precisely balanced along the line $K_2=-2K_1$.
As a check of the crossing between the two curves, we have represented 
the specific heat $C_{v,4}(K_1,-2K_1)$ for $L=4$ in Fig.\koneto. We see clearly the
four critical points, two of which lie on the continuous
transition curve, and two on the first order one.

\fig{The extrapolated critical lines of the short and long rigid
edge model in the plane $(K_1,K_2)$. The solid line curve represents the
first order (ferromagnetic) critical line, involving in particular 
values $K_1>0$ and 
$K_2>0$. Large positive $K_1$'s and $K_2$'s correspond to ferromagnetic order,
i.e. the completely flat phase. 
The dashed line curve represents the continuous (anti-ferromagnetic)
Ising-like transition line, involving in particular values 
$K_1<0$ and $K_2<0$. Large negative $K_1$'s and $K_2$'s correspond to 
anti-ferromagnetic order, hence a completely folded phase.}{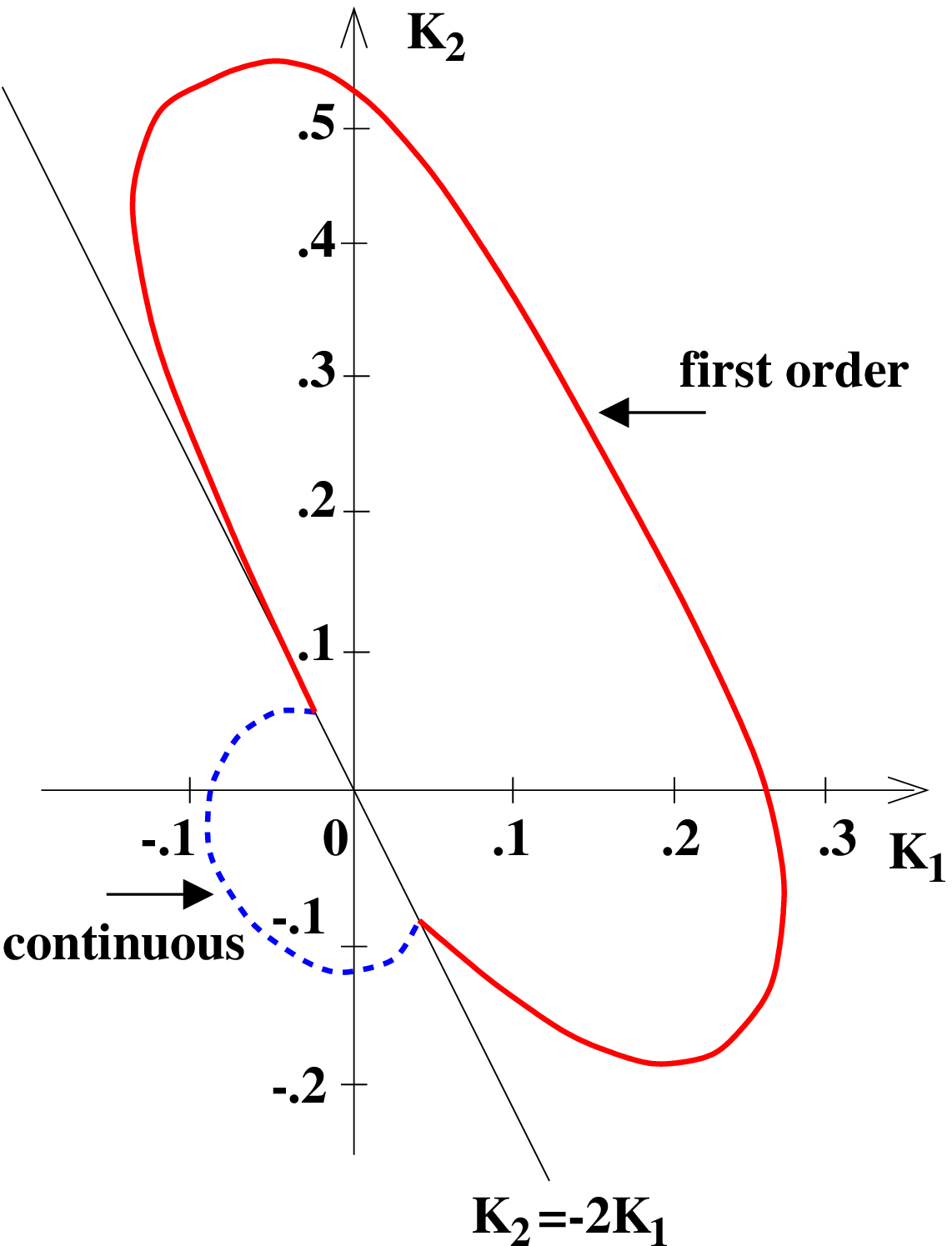}{6.cm}
\figlabel\phase

We are naturally led to conjecture the thermodynamic picture of Fig.\phase\
when $L\to \infty$, with two transition curves, continuous (anti-ferromagnetic)
in dashed line, and first order (ferromagnetic) in solid line. 
Outside of the solid curve, and above the line $K_2=-2K_1$, the model
is always in the completely flat phase with magnetization
$M=1$ if $h\to 0^+$ and $M=-1$ if $h\to 0^-$. Outside of the dashed curve,
and below the line $K_2=-2K_1$, the model is in a completely folded phase,
with possibly some unfolded loops of edges. 
The staggered (anti-ferromagnetic) magnetization $M_{st}$ corresponding to a
staggered magnetic field $h_{st}$ (taking alternating values $\pm h_{st}$ 
on adjacent triangles) is a function of the $K$'s tending to $1$ 
if $h_{st}\to 0^+$
and $-1$ if $h_{st}\to 0^-$ for large negative $K$'s.
The dashed curve corresponds to the vanishing of the staggered magnetization,
which becomes identically zero inside the curve.
So the region inside both curves corresponds to both magnetizations 
equal to zero: we can think of it as a partially folded (or unfolded) phase.

\fig{A rough sketch of the surface $h=h_c(K_1,K_2)$ for the critical
magnetic field separating the flat and folded phases of the
membrane. This surface coincides with the plane $h=0$ outside of
the transition curve and in the domain $K_2>-2K_1$  
Note the conic singularity at the point $(0,0,h=s_{SD})$,
and the jump singularities along the plane $K_2=-2K_1$.}{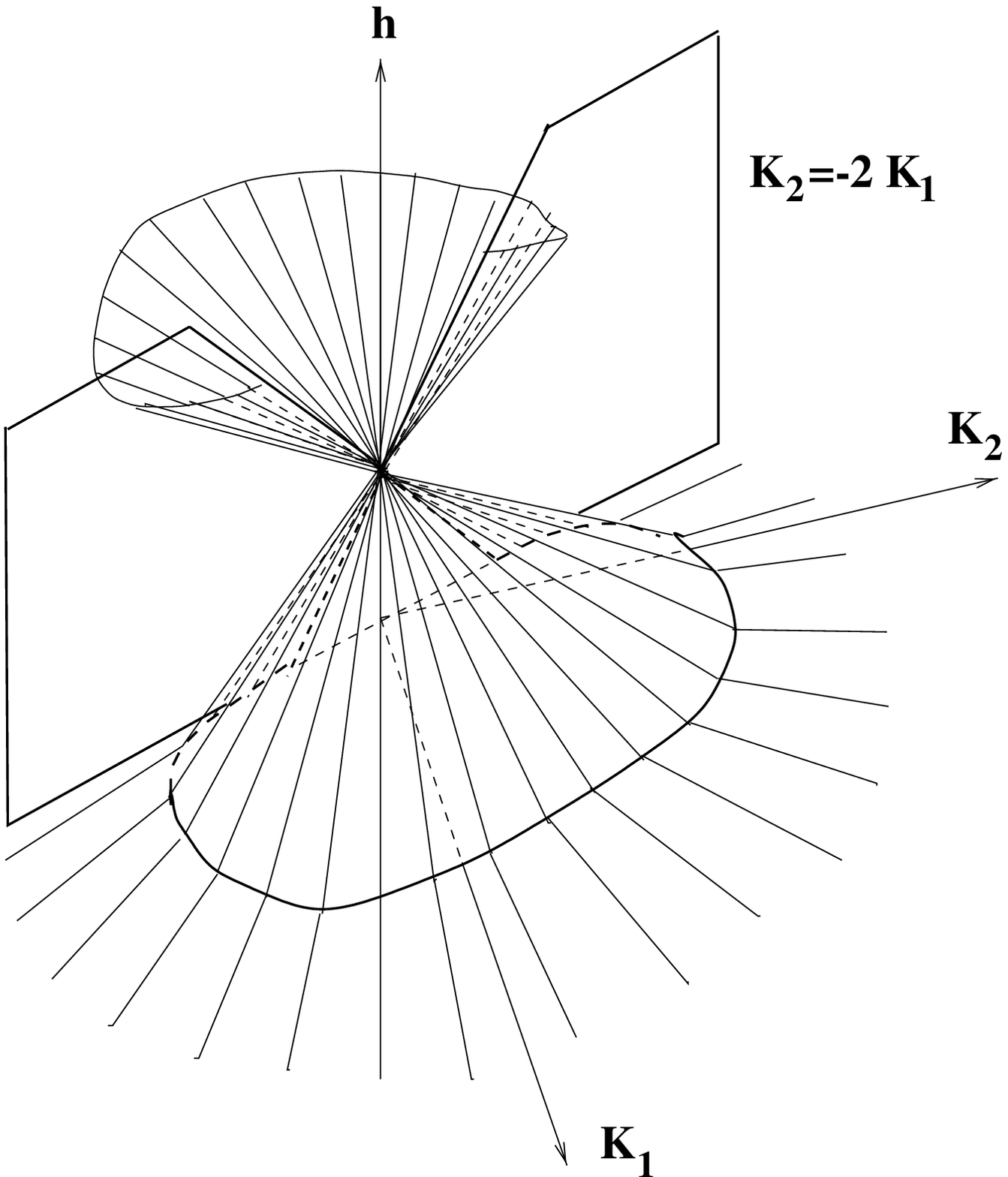}{8.cm}
\figlabel\hglo

To fully illustrate the ferromagnetic transition, we have represented
schematically (most of the curves are represented as lines) 
in Fig.\hglo\ the surface $h=h_c(K_1,K_2)$ of the critical magnetic field
beyond which the membrane is in the completely flat phase, and below
which it is folded. This surface is singular along
the plane $K_2=-2K_1$: $h(K_1,K_2)$ has a jump discontinuity from one
of its sides to the other, except at $K_1=K_2=0$.
The corresponding point $h=s_{SD}=h_c(0,0)$ is a conic singularity
of the surface. 

\bigskip

\leftline{\bf Acknowledgements}

This work was partially supported by NSF grant PHY-9722060.

\vskip 2.cm

\listrefs
%\refslist
\end